\documentclass{aa}
\usepackage{epsfig,graphics}
\begin{document}


%
   \title{First optical identification of a supersoft X-ray source in M31}
  \authorrunning{Supersoft source in M31}

 \author{P. Nedialkov \inst{1,} \inst{2,} \inst{3}, M. Orio \inst{4,} \inst{5}, K. Birkle \inst{6},
C. Conselice \inst{7}, M. Della Valle \inst{8}, J. Greiner \inst{9}, 
E. Magnier \inst{10,} \inst{11}, N. A. Tikhonov \inst{12} } 
   \offprints{M. Orio}

   \institute{Department of Astronomy, Sofia University, Bulgaria
              \and Isaac Newton Institute of Chile, Bulgarian branch
              \and   japet@phys.uni-sofia.bg
              \and 
              INAF-Osservatorio Astronomico di Torino, Strada Osservatorio,
              20, I10025 Pino Torinese (TO), Italy
              \and
              Department of Astronomy, 474 N. Charter Str., Madison WI
              53706, USA
              \and
              Max-Planck-Institut f\"ur Astronomie, K\"onigstuhl, D-69117 Heidelberg, Germany 
              \and
              Department of Astronomy, Caltech, MC 105-24, Pasadena, CA 91125, USA
              \and
              Osservatorio Astronomico di Arcetri, Florence, Italy 
              \and
              Astrophysikalisches Institut, 14482 Potsdam, FRG
              \and
              University of Hawaii, Institute for Astronomy, 2680 Woodlawn Dr., Honolulu, HI 96822, USA
              \and
              Canada France Hawaii Corporation, PO Box 1597, Kamuela, HI 96743, USA 
      \and
      Special Astrophysical Observatory of the Russian Academy of Sciences,
 Nizhnyj Arkhyz,  Russia}
\date{Received ; accepted }



\abstract
{We propose the first association of an optical counterpart with
a luminous supersoft X-ray source in M31, RX J0044.0+4118, 
observed with {\sl ROSAT} in July 1991. The {\sl PSPC} position is 
at 1.6\arcsec angular distance from 
a candidate nova in outburst in September 
of 1990. This is interesting because the incidence of classical novae 
among supersoft X-ray sources is an open question.  
The proposed optical counterpart was measured at R$\simeq$17.7
in September of 1990, and it had faded to
R$>$19.2 when it was observed again after 70 days.  
 The light curve was too sparsely monitored for definite conclusions on the
speed class of the nova.
No other variable objects with V$<$23.5 were found in the  {\sl ROSAT}
 spatial
error box.  We evaluate that the probability that a classical or recurrent nova 
was in outburst in the {\sl ROSAT} error box in the few 
years preceding the observation  is very small, so the proposed identification
is meaningful. We show evidence that
the associated supersoft X-ray source turned off in the third year after the outburst.
}
\maketitle
\section{Introduction}

Supersoft X-ray sources (SuSo) are among the most intriguing objects
discovered in the past few years. Their X-ray luminosity
is in the range 10$^{36}$-10$^{38}$ erg s$^{-1}$,
a non-negligible fraction of the
Eddington luminosity of a 1 M$_\odot$ star 
(about 10$^{38}$ erg s$^{-1}$). The spectrum is blackbody-like,
with an effective temperature in the range 20-80 eV. 
The first supersoft X-ray sources were discovered with {\it Einstein}
in the Magellanic Clouds (Long et al. 1981; Wang et al. 1991) and later defined  
as a {\it whole new class of objects} thanks
to new {\sl ROSAT PSPC} discoveries in the Galaxy
and local group galaxies (see Kahabka
\& van den Heuvel 1997). A wide range of sources have been
classified as  SuSo  and they all seem to have one
important feature in common: a white dwarf burning hydrogen
in a shell. This class comprises 
the hottest PG1159 objects (e.g. Cowley et al. 1995),
rare planetary nebulae with low absorption of the surrounding
nebula, and a large number of
binary sources. Among the latter we have symbiotic
stars, classical and recurrent novae, 
V Sge stars, and a new class of
close X-ray binaries (van den Heuvel et al. 1992; Greiner 2000 and references therein). 

The close binary 
SuSo sources are particularly interesting as candidate
progenitors of type Ia supernovae. 
Kahabka (1999) estimates that 20\% to 100\% of all type Ia supernovae in M31
have SuSo progenitors.
Sixteen supersoft X-ray sources (hereafter SuSo) were detected during the
deep {\sl ROSAT PSPC} pointings of M31 (Greiner 1996; Supper et al. 1997;
Greiner 2000). Eighteen  other {\sl ROSAT} sources
in the direction of M31 have been proposed to belong
to this class (Kahabka 1999).  Most fields were subsequently observed with
the {\sl ROSAT PSPC} and the positions are precise within
10-15 arcsec. It is crucial to determine the precise nature
of this population and whether it is associated with
M31.  Yet, only nine of the sources have been scheduled for Chandra observations,
which will yield arcsecond precision if the sources are not transient. 
Foreground sources
(especially cataclysmic variables of the AM Her type,
which can be as ``soft'' as SuSo
but at least two order of magnitudes less luminous) have been
excluded by Greiner et al. (1996). Some sources
may be 
background AGN or large-amplitude outbursts of supersoft X-ray emission
in active and non-active 
background galaxies, which have been interpreted as tidal disruption events
(Komossa \& Greiner 1999).

 However, the number
of fore- and background sources in the direction
of M31 is limited, and the majority of objects in the field
belong to the  M31 population (at least among the original 16 SuSo
identified in the deep pointings). 
SuSo physically belonging to M31 can be detected easily
due to the lack of strong interstellar  absorption, 
and to the high bolometric luminosity which peaks in the supersoft
X-ray range. 
Most other SuSo were detected in the Magellanic Clouds 
and not in the Galaxy (see Orio et al. 1994, 1997; 
Greiner 2000 and references therein).  
The only way to know whether
one specific source belongs to M31 is to    identify and study the
optical counterpart.

\section{Classical and recurrent novae in M31}

Extrapolating from V$\simeq$11
for the brightest SuSo in the Galaxy and an average V$\geq$17
in the Magellanic Clouds, we conclude that
SuSo which belong to M31 are not expected
to be brighter than optical magnitude V$\simeq$22 
 unless they are classical or symbiotic novae shortly after the
optical outburst. Novae in M31 are known to have
maximum magnitude in the range V=16-18 (e.g. Rosino 1964a and b). The largest
fraction of M31 novae peak around the low end
of this range, and  are mostly {\it slow} novae (see Capaccioli et al.
1989). As long as there is no precise Chandra position, we need to rely only on
characteristics like very ``blue'' colour indexes,
optical variability, and specially bright emission
lines to identify the counterparts (see the studies of sources 
in the Magellanic Clouds, like Orio et al., 1997). 

Classical and recurrent novae in outburst, because their
brightness at maximum singles them out, are the easiest 
super-soft X-ray sources to identify optically in M31.  However, up to now only
5 out of 120 known novae observed with {\sl ROSAT} and other 
X-ray observatories in the Galaxy and Magellanic Clouds have been observed to 
turn into SuSo
(Orio et al. 2001a). Out of 120 novae observed with {\sl ROSAT},
 44 had an outburst
in the previous 10 years. Even if a fraction
of Galactic novae might have been affected by huge 
interstellar absorption, and very few observations
were repeated, the present statistics indicate 
only $\approx$10\% of classical and recurrent novae turn
into super-soft X-ray sources for at least  few years
(see Orio et al. 2001a). 
If this is confirmed, we must conclude that, in most nova systems,
the white
dwarf does not  retain part of the accreted mass after each outburst,
 a condition necessary for it to become a  
type Ia supernova (at least for most
type Ia SNe progenitor scenarii, see Yungelson \& Livio
1998).  What is special and distinct about
this sub-class of novae which do not expel all the accreted mass? 
In order to understand this issue,
it is very important to broaden the statistics by studying the
large number of M31 objects and identifying the novae among them.

The observed classical nova rate in M31 is 20 year$^{-1}$. 
Once it is corrected for extinction and other observational bias
the ``true'' nova rate turns out to be 
about 29$\pm$4 year$^{-1}$ (Capaccioli et al. 1989).
Since the 34 SuSo candidates were observed
in deep pointings that lasted for $\approx$2 years, we
might have had about 60 novae in that time span, most
of them concentrated in the bulge. However,
the detection efficiency of SuSo in
the bulge is reduced compared with the disk since source crowding makes
it difficult to extract the X-ray spectrum and determine the SuSo nature
of these sources.  This bias away from the bulge will further reduce the
chance of detecting a SuSo-nova in M31. 
As a matter of fact,  among 10 optically identified SuSo  in the
Magellanic Clouds, only one was a classical nova and another
one a symbiotic nova. The detection rate of
SuSo-novae in M31 will necessarily be lower than in the Magellanic
Clouds  which do not suffer the source crowding bias.
Therefore, we do not expect  a very
large fraction of SuSo-novae in the M31 sample,  but 
we definitely expect a few. The XMM transient XMMU J004319.4+411759
 has tentatively
been proposed to be a classical nova (Trudolyubov et al. 2001) even if
the optical nova counterpart has not been found. Along
with persistent SuSo, several transient sources exist both
in the Magellanic Clouds and M31 (e.g. White et al. 1995, Orio et al.
1997) but they may not all be novae, specially considering that
the Magellanic Clouds have been constantly monitored to discover
nova outbursts in the last 50 years (see Orio et al. 1997 and references therein).   
According to Yungelson et al. (1996) and Kahabka
\& Ergma (1997) other transient SuSo should exist which
are not classical or recurrent novae.
 
\section{The ROSAT X-ray observations of RX J0044.0+4118}

In this article we propose a classical nova as the optical counterpart of a 
SuSo discovered with {\sl ROSAT}.
RX J0044.0+4118 (or 2RXP J004404.8+411820) was detected in a {\sl PSPC} exposure of M31
done in July 1991 for about
25 ksec (Supper et al. 1997). The count rate we obtain, in agreement
with Supper et al. (1997), is    0.00252$\pm$0.00039 cts s$^{-1}$.
The software we used for the data analysis is {\sl EXSAS}
 (see Zimmermann et al. 1994).
The field was observed again for very short pointings with
the {\sl ROSAT PSPC} in July 1992, January 1993 and August 1993. Superimposing
the images Supper (2001, private communication)
obtains a 2 $\sigma$ upper limit $\leq$0.0008 cts s$^{-1}$.
Therefore, the flux had dropped by at least a factor of $\approx$4 during
the third year post--outburst.
In {\sl HRI} images, including very deep exposures done in 1996,  the source was not
detected. The best 3$\sigma$ {\sl HRI} upper limits
of 1996 are $\leq$0.0003 counts s$^{-1}$,
corresponding to approximately $\leq$0.0026 cts s$^{-1}$ in the {\sl PSPC}.
The rapid turn-off (t$_{\rm SuSo}\leq$ 3 years) in X-rays 
is not unusual for a post-outburst
classical nova, as we described in Section 2. According to Kato (1997)
10 months$\leq$t$_{\rm SuSo} \leq$ 3 years translate into
constraints into the nova white dwarf mass: 
0.8 M$_\odot \leq$ M$_\odot \leq$1.05 M$_\odot$ for low metallicity.

\section{Discovery of an object in outburst in the field of RX J0044.0+4118} 

Members of this team (see Orio et al. 2002
for preliminary results) conduct a project using the WIYN 3.5m telescope
located at Kitt Peak, Arizona, to image the fields
of the SuSo in M31. We obtain magnitudes
of objects in the spatial error boxes of {\sl ROSAT},
in different colours and usually 
with completeness limit $\simeq$23-24 in
different filters.  We observed the candidate SuSo in different bands,
and as first and most important step we identify the objects
with U and B excess and those that are variable. 

The   position of RX J0044.0+4118, $\alpha_{(2000)}$=0h 44m 4.76s
and $\delta_{(2000)}$=+41$^{\rm o}$ 18$\arcmin$ 20,2$\arcsec$ in the {\sl ROSAT}
``{\it rospspctotal}'' catalog of {\sl PSPC} sources
in HEASARC\footnote{High Energy Astrophysics Research Centre
at NASA-Goddard Space Flight Centre},  
is only 1.6  arcsec distant from an object detected 
around 18th magnitude during  observations of M31 performed in September
1990.  Photometry of this field was published
by Magnier et al. (1992) in the catalog of M31, 
obtained with CCD photometry done at the Mc-Graw Hill 1.3m telescope of
MIT at Kitt Peak, and later in an article 
by Nedialkov et al. (1996, see Tab.\,1), based on plates taken   
at the Special Astrophysical Observatory (SAO) of the Russian Academy of
Sciences 1m telescope. By sheer chance, the dates of observation 
of these fields in 1990 were very close:
1990 September 17, September 19 (Magnier et al. 1992) and September
21 (Nedialkov et al. 1996).  
We note that the object is No. 24 in Nedialkov et al. 1996, and the
position given in that paper has been revised here. 
Surprisingly, this object was not detected using the WIYN 3.5m telescope
in 1998 and 2000 (see Tab.\,2). We joined efforts with different
groups who have been observing M31 in the last 25 years and 
reviewed the observations prior to September 1990. All the 
results are reported in Tab.\,1 and Tab.\,2 (respectively, before and during,
and after outburst). 

The  limiting
magnitude was R$>$20.4 in a plate obtained with the MPI Calar Alto 0.8m
telescope on 1990 August 28 (see Tab.\,1), only 19 days before the
outburst detection.  The completeness limit of the
photometry in September 2000 was V$>$23.5. The 
maximum luminosity (equivalent to M$_{pg}\sim$6.5), the range of luminosity 
(rise by $\Delta$R$\geq$2.9 mag in about 3 weeks, 
decay by $\Delta$V=0.79 mag in less than 2 days and $\Delta$R$>$2.5 mag in 70
 days, and $\Delta$V$>$5.3 mag  in a few years) suggest a classical nova in M31,
 which     
must have had an outburst on 1990, between August 30 and September 17. 
We found that while the V magnitude was already decreasing in the 2 days following
the first detection on 
September 17. However, perhaps the object was becaming 
slightly redder (it could have been due to dust formation) and definitely 
bluer (which is indeed expected for
classical novae shortly after the outburst). However, the 
data in Tab.\,1 are too sparse for definite conclusions.
We note that the ``fragment''
of light curve we have is consistent with two different
types of light curves.  If we caught the nova at maximum,
the few points we have could fall on the light curve of a slow nova,
perhaps a nova with more than one peak (see novae
No. 69 and No. 80 of Rosino, 1973). Such
 a nova would probably belong to the thick disk
or bulge population. On the other hand, if 
 instead we missed the maximum, and the maximum magnitude
was V$\simeq$16.5, the nova could be a fast one, like
No. 52 of Rosino (1973), and belong to the disk population.
The short turn--off time in X-rays seems to point at this
second possibility as more likely. 
The position of our nova candidate 
is $\alpha_{(2000)}$=0h 44m 04.71s $\delta_{(2000)}$=41$^{\rm o}$
18$\arcmin$ 21.7$\arcsec$.  Six different telescopes yielded 
the results in Tab.\,1 and Tab.\,2, in which the magnitudes of a 
nearby star (No. 5) are also reported: the 1.2m and 0.8 m of
 the Max Planck Institute
at Calar Alto, Spain, the 1m telescope of the SAO (Russian Academy of
Sciences), the  2m telescope of the BAS
(Bulgarian Academy of Sciences), the  WIYN 3.5m telescope at Kitt Peak
in 1998 August and in 2000 September, 
and finally the 6m telescope of the SAO in 2000 October.
The WIYN observations were done  with a single CCD 
in August 1998 and with the   Mini-Mosaic Imager (4 frames in the mosaic,
obtained with
2 CCDs of 4096x2048 pixels, each read by two amplifiers) in September 2000.
An optical image obtained in September 2000
and the comparison with the plate taken during to the outburst is shown in 
Fig. 1. The decline in luminosity of classical novae in M31 is on average slower than for
Galactic novae, and a decrease by more than 2.5 magnitudes in 70 days, 
is only observed in about half of the M31 novae.

As it can be seen in Tab.\,1 and Tab.\,2, we did not detect significant
 variability of 
star No. 5 and the apparent colour indexes do not suggest a very hot star
(i.e. a competing ``blue'' candidate for the optical counterpart). 
Objects with larger magnitude than the  completeness limit
can still be detected in the different images
(e.g. several stars at V=24-24.5 in the WIYN exposures
of September 2000), but the sample is not significant below the  completeness
limit which we 
adopt as a  $\simeq$3$\sigma$ upper limit. 
In Tab.\,3 we give the BVRI photometry for  stars within
the 15\arcsec radius error circle corresponding to the 3$\sigma$ {\sl ROSAT}
 position 
of RX J0044.0+4118, detected in at least two different bands 
with V$\leq$23.5. 

Comparing the WIYN exposures taken in two consecutive
nights in 2000, in two consecutive nights in 1998 and also
with Magnier et al. (1992) and Nedialkov et al. (1996), 
we did not find other  variable objects in the spatial error box
of the {\sl PSPC} over time scales of a day or years.  There are 13 other
``blue'' stars in Tab.\,3 (B-V$\leq$0.4) 
which we  consider OB-candidates.  
For the stars in Tab.\,3 detected both in the R and I
passband, we plotted in Fig 2 a colour-magnitude diagram.
We also plotted the Zero Age main Sequence and the evolutionary
tracks (from Pols et al., 1998, transformed into V versus V--I using the
Basel interactive server, Lejeune et al. 1997), for m=2 and 4 M$_\odot$,
adopting the average foreground extinction towards M31 E(B--V)=0.062
(see Schlegel et al. 1998).
In Fig. 3 we plotted instead a  
colour-colour diagram for the stars with B, V and I detections.
In both figures we show the position of 
star No. 2, our candidate nova, when it was observed in September of 1990. 
Because most stars detected in the three colours are very faint, the points are spread in Fig.3, with large errorbars. Fourteen stars
in the Figure deviate from the expected range of extinction law. However, this
could be due to blending with neighbouring stellar images (low mass OB stars
on the main sequence and AGB stars). 

It can
be clearly seen that the nova candidate stands out in the upper left corner
in the colour-magnitude plot (Fig. 2) and that it had typical colours of
a not heavily absorbed nova (Fig. 3; note that we consider
E(B--V)=0.16 intrinsic to M31).  
The variability time scales we sampled for the putative
nova  are several years  by comparisons
with different telescopes; with WIYN alone we sampled  
$\simeq$2 years between 1998 and 2000, 1 day, and few minutes over
a 2 hour period. The only detection and the only
outburst we discovered are the ones of 1990 September 17-21.   

Finally, we would like to add a few remarks about the stellar
population  within the 15$\arcsec$ 
error circle. Fig 2 shows that it is dominated by an old disk population 
(71 of total 86) of AGB stars, clumped 
around the values (R--I)=1.2 and I=20.5 and belonging to the
``Tip of the AGB'' (see HST photometry by Sarajedini \& Duyne 2001).
The interval between the evolutionary tracks (Pols et. al, 1998)
for 2 and 4 solar masses and 0.9$<$(R--I)$<$1.6 corresponds to stellar
ages in the range of 0.2--1.5 Gyr. 
  Most of the other 15 stars (excluding the nova) could be
OB stars according to different colour criteria (such
as (B--V)$<$0.4, for example). The {\sl ROSAT} error circle is located out
of the boundaries of OB associations in M31 and there are no
indications of recent star formation.
Moreover, we also obtained an H$\alpha$ image with the WIYN
telescope in 2000 September, and failed to detect
any diffuse nebulosity, which seems to exclude
the presence of an HII region  other than No. 23 of Tab.\,3, 
 k282. Star No. 5, the brightest one (V=19.2) in the
field, could be a reddened OB supergiant although
neither apparent colour indexes nor dereddening according to
the classical extinction law  (R$_{\rm V}$=3.1) suggest a very hot
star.  Examining  a 25 times larger area centered at the X-ray
source position and comparing it with the possible foreground
contamination (negligible for a circle with a radius of 15 $\arcsec$)
leads to  only a 10\% probability that star No. 5 belongs
 to the Galaxy. It is more  likely to be an OB star.
Star No. 53 it is the brightest main sequence OB candidate at V=20.6.
Its absolute magnitude  could be as large
as M$_{\rm V}$-5.8 after dereddening with 
A$_{\rm V}$=2 mag. Some of the other blue star candidates are
too faint to be detected in shorter wavelengths and they are not
shown in Fig. 2.

\section{The probability of a random coincidence}

Although probably only 10\% of all classical and recurrent novae
turn into SuSos, and our colour index and variability
study with the WIYN can detect only up to $\approx$50\%
of all possible SuSo counterparts (we estimate the rest to be at V$\leq$24.5 ) we
are able to show that the possibility of a random spatial
coincidence between the proposed nova and RX J0044.0+4118
is {\it not} high.
About 85\% of
the novae discovered in M31 are located within 12 arcmin
from the center of M31. Their number decreases with distance from the center. 
RX J0044.0+4118 is at a distance 15.29 arcmin from the center of
M31, which we consider to be
$\alpha_{(2000)}$=0h 42m 44.16s and $\delta_{(2000)}$=41$^{\it o}$
16$\arcmin$ 08.6$\arcsec$.  From the data published by 
Capaccioli et al. (1989), we estimate that the probability that one of the
novae in outburst and observed in one year
falls in a 15 arcsec spatial error box at a distance
15.29 arcmin from the center of M31 is only  $\approx 1.5 \times 10^{-4}$, 
or in other words in 10$^5$ years we expect only 15  novae in
the error box. Even smaller is the probability of random 
coincidence with a recurrent nova. Della
Valle and Livio (1996) have shown that the ratio
between the rate of outbursts of recurrent and classical novae is in
the range 0.1--0.3, thus the probability to run into a non-physical
coincidence is not larger than $0.05\%$
Assuming that the constant bolometric luminosity
phase lasts for the upper limit we have on the life
of RX J0044.0+4118 as SuSo, 3 years, we obtain a 0.05\% probability to detect a
nova within a 15 arcsec spatial error box of a SuSo, 
 without being physically associated with it.
We point out that this is even a very conservative conclusion, assuming
that {\sl all} the sources discovered in the M31
direction belong to the galaxy and {\it all} novae turn into SuSo. 
However, since  observational
estimates suggest that only
about 10\% of novae are detectable for few years in supersoft X-rays,
 it is likely that the probability
of a random coincidence is even smaller than quoted above, 
probably of the order of $\sim$0.01\%. 
%
\begin{table}
\caption[Table]{Results of observations in different
nights and years: date, telescope,  band of the filter, 
absolute magnitude or upper limit of the nova candidate and of star no. 5,
the brightest star in the {\sl ROSAT} error circle.
SAO= Special Astrophysical Observatory of the Russian Academy of Sciences, BAS=Bulgarian
Academy of Sciences, MGH=McGraw-Hill, MPI=Max Planck Institut (FRG). $\sigma$
is the error on the magnitude or colour index of the previous column.
The asterisk near ``0.8 MPI'' indicates that
an objective prism was mounted. }
\label{Table}
\[
\begin{array}{l c c c c l}    
\hline 
$date$ & $telescope$ & $band$ &  $mag(n)$ & $mag(5)$ & \\  
\hline
1976/09/15 & $1.2m MPI$ & $B$ & >19.5  & >19.5  \\ 
1976/11/15 & $1.2m MPI$ & $B$ & >20.8  & 20.2\pm0.1 \\
1976/11/17 & $1.2m MPI$ & $B$ & >22.1  & 20.2\pm0.1 \\
1976/11/20 & $1.2m MPI$ & $B$ & >22.1  & 20.1\pm0.1 \\
1977/08/11 & $1.2m MPI$ & $V$ & >20.7  & 19.5\pm0.1 \\ 
1977/08/12 & $1.2m MPI$ & $B$ & >20.1  & 20.1\pm0.1 \\
1977/10/04 & $1.2m MPI$ & $B$ & >21.0  & 20.2\pm0.1 \\
1980/10/15  & $BAS$  & $B$  &  > 22.0  & 20.0\pm0.1 \\    
1981/10/03  & $BAS$  & $B$  &  > 21.2  & 20.1\pm0.2 \\     
1981/10/04  & $BAS$  & $B$  &  > 21.8  & 20.2\pm0.2 \\      
1981/10/05  & $BAS$  & $B$  &  > 21.2  & 20.0\pm0.2   \\
1982/07/22 & $1.2m MPI$ & $B$ & >20.9  & 20.2\pm0.1 \\
1982/11/09  & $BAS$  & $B$  &  > 22.0  & 20.5\pm0.1   \\
1982/11/13  & $BAS$  & $B$  &  > 22.0  & 20.0\pm0.2   \\
1982/11/22  & $BAS$  & $B$  &  > 21.5  & 20.2\pm0.2 \\    
1982/11/23  & $BAS$  & $B$  &  > 21.5  & 20.0\pm0.2 \\
1982/12/05 & $0.8m MPI$ & $B$ & >21.3  & 20.0\pm0.1 \\
1983/01/13  & $BAS$  & $B$  &  > 21.5  & 20.1\pm0.3 \\
1983/08/11  & $BAS$  & $B$  &  > 21.5  & 20.1\pm0.2 \\
1983/10/15  & $BAS$  & $B$  &  > 21.0  & 20.1\pm0.2 \\
1983/10/16  & $BAS$  & $B$  &  > 21.0  & 20.1\pm0.2 \\
1983/12/23 & $0.8m MPI$ & $R$ & >19.7  & 19.0\pm0.2 \\
1984/09/05 & $0.8m MPI$ & $R$ & >19.0  & >19.0  \\
1984/10/24  & $BAS$  & $U$  &  > 21.0  & 20.4\pm0.4 \\
1985/07/18  & $BAS$  & $B$  &  > 21.5  & 20.3\pm0.2 \\
1986/10/03  & $BAS$  & $V$  &  > 20.5  & 19.1\pm0.2 \\
1986/11/02  & $BAS$  & $V$  &  > 21.2  & 19.1\pm0.1 \\
1987/10/27 & $0.8m MPI*$ & $R$ & >17.5  &  >17.5 \\
1990/06/29 & $0.8m MPI$ & $R$ & >20.5  & 19.0\pm0.2 \\
1990/07/04 & $0.8m MPI$ & $R$ & >20.5  & 19.0\pm0.2 \\
1990/08/02 & $0.8m MPI$ & $R$ & >20.4  & 19.0\pm0.2 \\
1990/08/17 & $0.8m MPI$ & $R$ & >19.5  & 18.9\pm0.2 \\
1990/08/28 & $0.8m MPI$ & $R$ & >20.4  & 19.0\pm0.2 \\
1990/09/17.44  & $MGH$        & $R$  & 17.711\pm0.025 &  18.744\pm0.004\\
1990/09/17.45  & $MGH$        & $I$  & 17.372\pm0.025 & 18.437\pm0.037 \\
1990/09/17.45  & $MGH$        & $V$  & 18.054\pm0.025 & 19.207\pm0.027\\
1990/09/17.45  & $MGH$        & $B$  & 18.428\pm0.025 & 19.998\pm0.034 \\
1990/09/19.35  & $MGH$        & $V$  & 18.845\pm0.025 &   \\ 
1990/09/19.36  & $MGH$        & $I$  & 18.390\pm0.025 &   \\
1990/09/20.99  & $0.8m MPI$       & $R$  & 17.65\pm0.08   &  18.83\pm0.10 \\
1990/09/22.03  & $1m SAO$  & $B$  &   18.2\pm0.1 &   \\              
1990/12/12  & $MPI$     & $R$  &   \geq 19.20  & 18.73\pm0.09 \\
\hline
	 \end{array}
\]
\end{table}
\begin{table}
\caption[Table]{Same as in Table 1 for observations
done since the beginning of 1991. The asterisk near ``0.8m MPI''
indicates that an objective prism was mounted.} 
\[
\begin{array}{l c c c c l}
\hline
$date$ & $telescope$ & $band$ &  $mag(n)$ & $mag(5)$ & \\
\hline
1991/01/03  & $0.8m MPI$ & $R$ & >19.3 & 18.72\pm0.07 \\
1991/01/18  & $0.8m MPI*$ & $R$ & >19.8 & 18.8\pm0.2 \\
1991/06/19  & $0.8m MPI$ & $B$ & >20.9 & 20.0\pm0.2 \\
1991/06/21  & $0.8m MPI$ & $B$ & >21.1 & 20.2\pm0.2 \\
1991/08/09  & $0.8m MPI$ & $B$ & >19.0 & 18.7\pm0.2 \\
1991/10/01  & $1m SAO$  & $B$  &  > 20.5  & 20.1\pm0.3  
\\
1991/10/06  & $1m SAO$  & $B$  &  > 22.0  & 20.2\pm0.2 
\\
1991/10/07  & $1m SAO$  & $B$  &  > 22.0  & 20.2\pm0.2  
\\
1991/10/09  & $1m SAO$  & $B$  &  > 21.4  & 20.4\pm0.2 
\\
1991/10/12  & $1m SAO$  & $B$  &  > 22.0  & 20.0\pm0.1  
\\
1991/10/29  & $0.8m MPI$       & $R$  & >20.5         & 18.8\pm0.2 \\
1991/11/11  & $0.8m MPI$       & $R$  & >20.5         & 18.8\pm0.2 \\
1991/12/12  & $0.8m MPI$       & $R$  & >19.3         & 18.9\pm0.1 \\
1992/01/27  & $1m SAO$  & $B$  &  > 21.5  & 20.3\pm0.2  \\
1992/01/28  & $1m SAO$  & $B$  &  > 21.0  & 20.0\pm0.2  
\\
1992/02/01  & $0.8m MPI$       & $R$  & >19.3         & 18.9\pm0.1 \\
1992/07/28  & $1m SAO$  & $B$  &  > 21.2  & 20.4\pm0.2  
\\
1992/07/29  & $1m SAO$  & $B$  &  > 21.4  & 20.0\pm0.2 
\\

1992/08/03  & $1m SAO$  & $B$  &  > 21.5  & 20.4\pm0.2  
\\              
1992/08/05  & $1m SAO$  & $B$  &  > 21.8  & 20.5\pm0.2  
\\              
1992/11/23  & $0.8m MPI*$       & $R$  & >18.8  & 18.8\pm0.1 \\
1992/11/25  & $0.8m MPI$       & $R$  & >20.4  & 18.8\pm0.1 \\
1993/01/24  & $0.8m MPI*$       & $R$  & >18.9  & 18.8\pm0.1 \\
1994/09/27  & $2m BAS$  & $B$  &  > 21.0  & 20.1\pm0.2  
\\              
1994/09/27  & $2m BAS$  & $B$  &  > 21.0  & 20.1\pm0.2  
\\                            
1994/09/28  & $2m BAS$  & $B$  &  > 21.0  & 20.0\pm0.2 
\\                            
1994/09/28  & $2m BAS$  & $B$  &  > 21.0  & 19.9\pm0.2   \\
1996/10/10  &  $0.8m MPI$       & $V$  & >20.5  & 19.2\pm0.1 \\
1996/12/02  &  $0.8m MPI$       & $V$  & >19.7  & 19.3\pm0.1 \\
1997/10/30  &  $0.8m MPI$       & $V$  & >20.0  & 19.4\pm0.1 \\
1998/08/17  & $WIYN$    & $R$  &  > 20.0  & 18.77\pm0.05  \\
1998/08/17  & $WIYN$    & $U$  &  > 20.2  & >20.2  \\
1999/01/19  &  $0.8m MPI$       & $V$  & >19.7  & 19.3\pm0.1 \\
2000/08/28  & $0.8m MPI$       & $V$  & >21.0  & 19.3\pm0.1 \\ 
2000/09/26  & $WIYN$    & $V$  &  > 23.5  & 19.18\pm0.02  \\
2000/09/27  & $WIYN$    & $V$  &  > 23.5  & 19.19\pm0.02  \\
2000/09/27  & $WIYN$    & $R$  &  > 22.0  & 18.77\pm0.03  \\
2000/10/02  & $6m SAO$  & $B$  &  > 22.5  & 20.08\pm0.02  
\\              
2000/10/02  & $6m SAO$  & $V$  &  > 23.0  & 19.20\pm0.01  
\\              
2000/10/02  & $6m SAO$  & $R$  &  > 22.5  & 18.74\pm0.01  
\\              
2000/10/02  & $6m SAO$  & $I$  &  > 21.5  & 18.27\pm0.01   \\ 
\hline
	 \end{array}
\]
\end{table}
\begin{figure}
\includegraphics[width=9cm,clip]{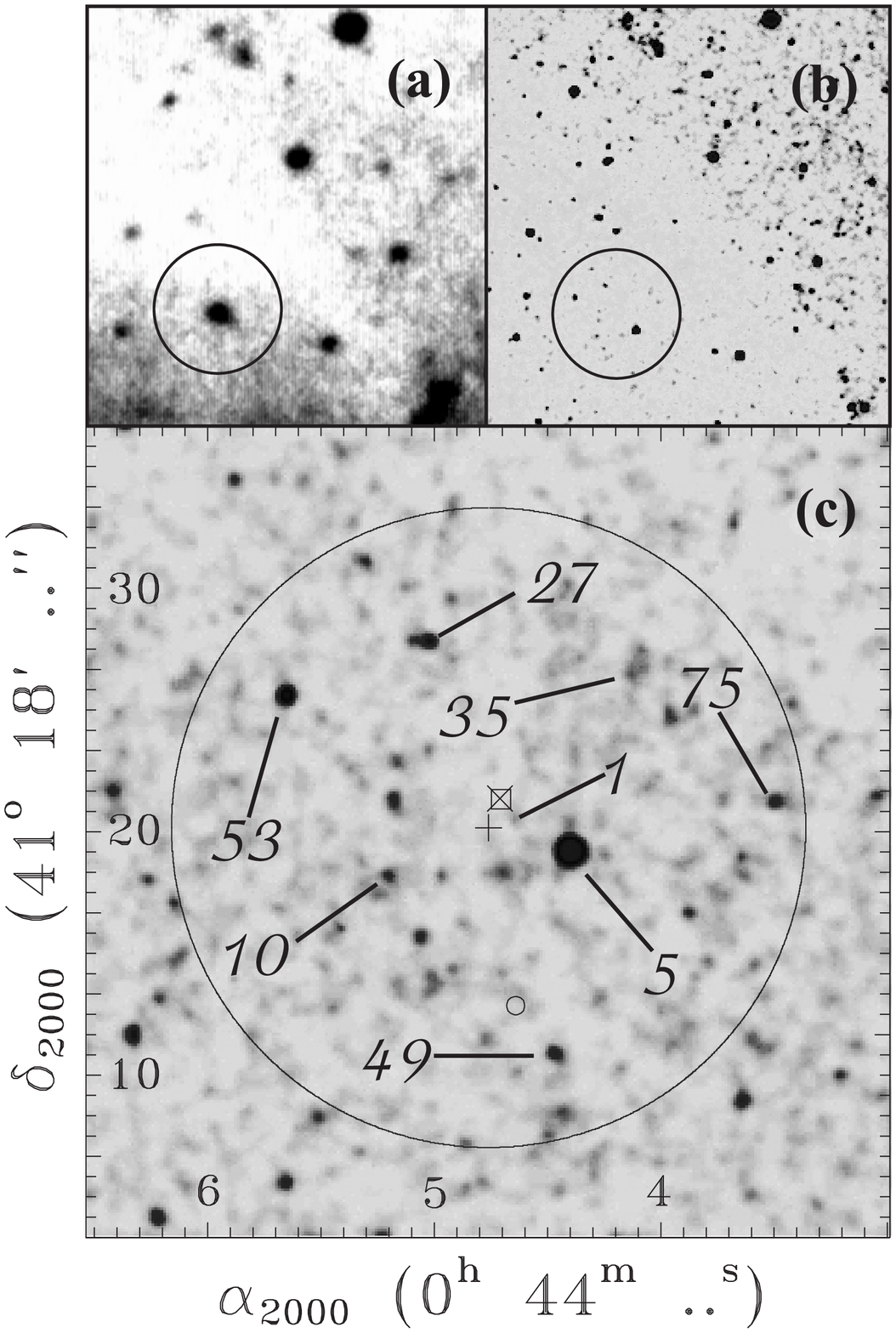}
\caption {A 15\arcsec error circle around the {\sl ROSAT} position of the 
supersoft X-ray source RX J0044.0+4118 is plotted here on different images. 
The small image on the left ({\it a}) was obtained in B band by Nedialkov et al. (1996)
on September 21 1990 and image {\it c} which   is a part of {\it b} was 
obtained in the V band with the WIYN 3.5m telescope in 2000 September. 
The position of the
putative nova is indicated by the open square (as measured
by Nedialkov et al., 1996) and by the ``X'' (as measured by Magnier et al., 1992) 
and in Tab.\,3 we report the photometric measurements obtained for all
the stars in the error circle. We show in {\it c} the brightest stars of Tab.\,3.}
\end{figure}    

\section{Conclusions}

The nature of the population of SuSo in M31 is 
still to determine, and in this article we ``broke the ice'' 
by proposing the identification of one of the sources,
RX J0044.0+4118, with
a candidate classical nova, which must have been in 
outburst 10 months before the {\sl ROSAT}
observations. Since the statistics on Galactic novae 
are still very incomplete, it is important to determine 
how frequently extra-galactic  novae turn into SuSo
and how they contribute to the SuSo population in near-by galaxies.
Moreover, this object would be only the sixth classical nova ever observed
to become a SuSo (four were 
observed in the galaxy andone in the LMC). Turn-off in supersoft X-rays occurred  
after about two years, which makes this object similar to the
Galactic nova V1974 Cyg (Krautter et al. 1996).
\begin{figure}
\includegraphics[width=8cm,clip]{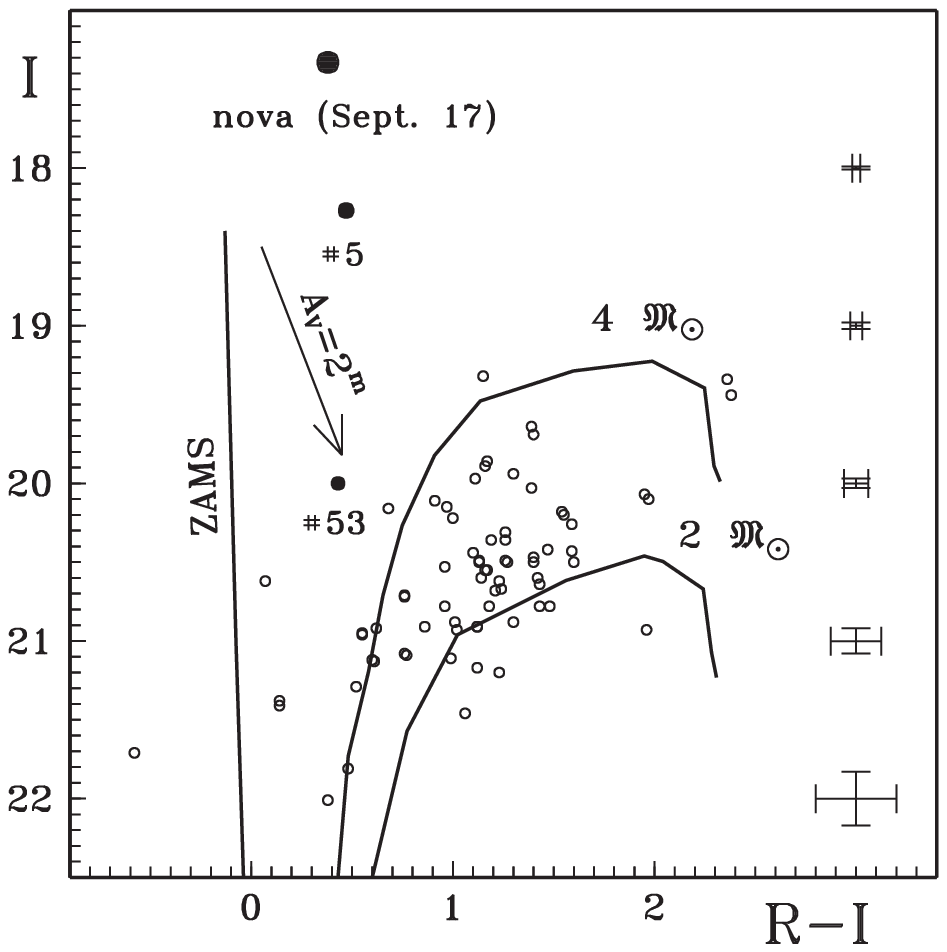}
\caption{Color-magnitude diagram for 77 stars 
with RI photometry in Tab.\,3. The approximate location of
ZAMS at the distance modulus (m-M)$_0$=24.47 of M31 (Stanek \& 
Garnavich, 1998)  is shown.  The evolutionary tracks 
 for m=2 and 4 M$_\odot$ 
 are also plotted (see text).
The arrow indicates
 the reddening vector corresponding to the classical extinction law
of Mathis (1990), and R$_{\rm V}$=3.1. The filled circles represent the
magnitudes and colours of the candidate nova on Sept. 17 1990, and of the
two brightest stars (No. 5 and 53) within the 15$\arcsec$ error circle from 
the nova. On the right the  average errors for different values of V 
are shown.}
\end{figure}
\begin{figure}
\includegraphics[width=8cm,clip]{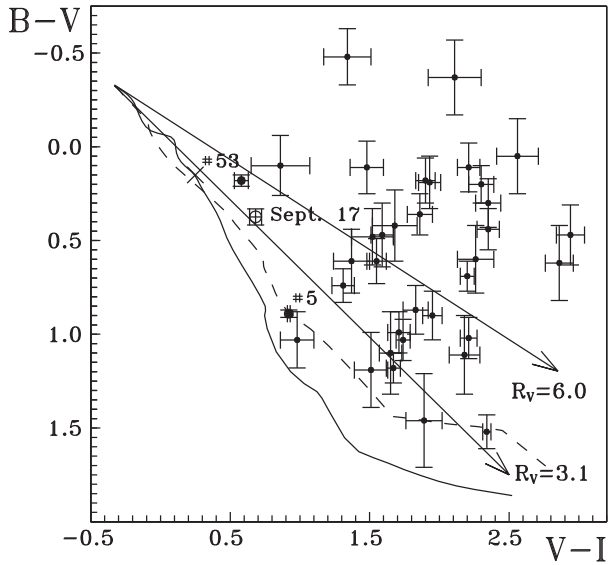}
\caption{Two-colour diagram for the 37 stars from Tab.\,3
 with BVI photometry. The solid line is the location of
supergiant stars and the long dashed line is the location of main
sequence stars (Bessel, 1990). The two arrows represent
the reddening vector typical for the ``diffuse'' (R$_{\rm V}$=3.1)
 and the ``dense''
(Rv=6.0) interstellar medium (Mathis 1990). The open circle indicates
the position of the candidate nova in this diagram on Sept. 17, 1990
and the two filled circles indicate the positions
of stars no. 5 and 53 in September of 2000. The mean locus for
Galactic novae at maximum, (B-V)$_o$=0.15 (Cohen, 1985),
 is marked with an ``X''.}
\end{figure}

  \begin{table*}
  \caption[Table]{Photometry obtained for all the stars
in the ROSAT-PSPC error circle of 15 arcsec. 
We give the V magnitude, the colour indexes V-R and R-I and their
1 $\sigma$ errors. The second column 
reports the angular distance in arcsec from the position of the supersoft
X-ray source, the last column reports the number in
the catalog by Magnier et al. (1996). the nova candidate is star no.2.}
\label{Table}
$$ 
 \begin{array}{l c c c c c c c c c c c c l}
\hline
 $no.$ &  $d$(\arcmin) & \alpha_{2000} & \delta_{2000} &  $V$   &\sigma&  $B-V or B$ & \sigma &  $V-R or R$ & \sigma & $R-I or I$ &
 \sigma &  $Mag.no.$  \\
\hline
 1 &  1.3 & $0 44 4.65$ & $ 41 18 20.7$ & 22.78 & 0.10 &  -   &  -   &  0.71 & 0.13 & 1.97& 0.10 &      \\
 2 &  1.6 &  $0 44 4.71$ & $41 18 21.7$ &  -    & -    & 18.20&  0.10&   -   & -    & -   & -    & 
265487 \\  
 3 &  3.7 & $0 44 5.00$ & $41 18 17.7$ &  -    & -    & -    & -   &  21.74 & 0.08 & 1.14& 0.08 &       \\
 4 &  3.7 & $0 44 5.09$ & $41 18 20.6$ & 22.57 & 0.07 & 0.11 & 0.13 &  1.02 & 0.09 & 1.19& 0.07 & 265443\\
 5 &  4.0 & $0 44 4.42$ & $41 18 19.0$ & 19.19 & 0.01 & 0.89 & 0.02 &  0.45 & 0.02 & 0.47& 0.01 & 265325\\
 6 &  4.3 & $0 44 4.59$ & $41 18 24.0$ & 22.74 & 0.11 & -0.35&  0.13 & 1.55 & 0.14 &  -   &  -   &      \\
 7 &  5.0 & $0 44 5.19$ & $41 18 21.5$ & 21.89 & 0.05 & 1.03 & 0.11 & 0.77 & 0.06 & 0.97 & 0.04 &  265443\\
 8 &  5.1 & $0 44 5.16$ & $41 18 17.9$ &  -    & -    & -    & -  &   21.62 & 0.08 &  1.26 & 0.08 &  265221\\
 9 &  5.6 & $0 44 5.16$ & $41 18 23.4$ & 22.07 & 0.06 & 1.02 & 0.11&  1.04 & 0.07  & 1.17 & 0.04 &  265565\\
10 &  5.7 & $0 44 5.22$ & $41 18 17.6$ & 21.90 & 0.07 & 0.47 & 0.17&  0.33 & 0.10  & 1.26 & 0.08 &  265221\\
11 &  5.9 & $0 44 4.45$ & $41 18 15.5$ & 22.83 & 0.15 &  -   &  -  &  0.31 & 0.19  & 1.06 & 0.14  &      \\
12 &  6.5 & $0 44 5.28$ & $41 18 17.5$ &  -    & -    & -    & -   & 21.50 & 0.08  & 0.55 & 0.08  &      \\
13 &  6.9 & $0 44 5.31$ & $41 18 17.3$ & 22.44 & 0.10 & 0.11 & 0.14&  0.93 & 0.11  & 0.55 & 0.08  &      \\
14 &  6.9 & $0 44 4.64$ & $41 18 13.5$ & -    & -    & -    & -   & 22.02 & 0.07  & 1.95 & 0.07  & 264909\\
15 &  7.2 & $0 44 5.35$ & $41 18 22.8$ & 22.45 & 0.12 & 0.61 & 0.17 & 0.61 & 0.14  & 0.76 & 0.09  &      \\
16 &  7.4 & $0 44 5.40$ & $41 18 18.5$ & 22.52 & 0.06 & 0.18 & 0.12 & 0.67 & 0.08  & 1.23 & 0.06  &      \\
17 &  7.4 & $0 44 4.35$ & $41 18 14.4$ & 22.58 & 0.06 &  -   &  -   & 0.47 & 0.09  & -    & -     &     \\
18 &  7.6 & $0 44 5.05$ & $41 18 27.0$ & 22.87 & 0.13 & 0.10 & 0.16 & 0.48 & 0.17  & 0.38 &0.20   &     \\
19 &  7.6 & $0 44 4.78$ & $41 18 12.6$ & 23.09 & 0.11 &  -   &  -   & 1.55 & 0.13  & 0.62 & 0.08  &      \\
20 &  7.6 & $0 44 4.68$ & $41 18 12.6$ & 23.04 & 0.18 & -0.37&  0.20&  1.09 & 0.19 & 1.02 & 0.08  &      \\
21 &  8.2 & $0 44 5.46$ & $41 18 18.2$ &  -    & -    & -    & -   & 21.91 & 0.06  & 1.24 & 0.07  &      \\
22 &  8.6 & $0 44 4.94$ & $41 18 28.5$ &  -    & -    & -    & -   & 21.74 & 0.06  & 0.96 & 0.08  &      \\
23 &  8.7 & $0 44 4.65$ & $41 18 11.6$ & 22.97 & 0.14 & 0.42 & 0.19 & 1.16 & 0.15  & 0.52 & 0.10  &
 $k282 (HII reg.)$ \\ 
24 &  8.8 & $0 44 5.46$ & $41 18 16.3$ & 22.73 & 0.10 & 1.11 & 0.21 & 1.02 & 0.11  & 1.16 & 0.06  &      \\
25 &  9.4 & $0 44 5.58$ & $41 18 18.0$ & 22.84 & 0.09 & 0.30 & 0.13 & 1.09 & 0.11  & 1.26 & 0.06  &      \\
26 &  9.5 & $0 44 5.48$ & $41 18 15.2$ &  -    & -    & -    & -   & 21.72 & 0.06  & 1.17 & 0.06  & 265076\\
27 &  9.8 & $0 44 5.07$ & $41 18 29.4$ & 22.13 & 0.10 &  -   &  -  &  0.59 & 0.11  & 1.10 & 0.04  &      \\
28 &  9.9 & $0 44 5.48$ & $41 18 14.5$ & 22.85 & 0.08 &  -   &  -  &  0.83 & 0.11  & 1.59 & 0.09  &      \\
29 &  9.9 & $0 44 4.29$ & $41 18 11.8$ & 22.22 & 0.06 & 0.74 & 0.09&  0.45 & 0.08  & 0.86 & 0.08  &      \\
30 &  9.9 & $0 44 5.51$ & $41 18 25.3$ & 23.15 & 0.35 &  -   &  -  &  2.31 & 0.35  & 0.68 & 0.04  & 265711\\
31 & 10.0 & $0 44 5.62$ & $41 18 22.6$ &  -    & -    & -    & -   & 22.89 & 0.15  & 1.96 & 0.15  &      \\
32 & 10.0 & $0 44 5.60$ & $41 18 17.0$ &  -    & -    & -    & -   & 21.77 & 0.07  & 1.27 & 0.07  &      \\
33 & 10.6 & $0 44 5.50$ & $41 18 13.5$ &  -    & -    & -    & -   & 22.10 & 0.08  & 1.60 & 0.09  &      \\
34 & 10.6 & $0 44 5.31$ & $41 18 28.7$ &  -    & -    & -    & -   & 22.18 & 0.09  & 1.30 & 0.09  &      \\
35 & 10.6 & $0 44 3.96$ & $41 18 25.7$ & 21.93 & 0.07 & 0.99 & 0.11 & 0.71 & 0.08  & 1.00 & 0.06  & 265764\\
36 & 10.6 & $0 44 5.65$ & $41 18 16.6$ & 22.73 & 0.10 &  -   &  -   & 1.03 & 0.12  & 2.36 & 0.07  & 265131\\
37 & 10.6 & $0 44 5.41$ & $41 18 12.5$ & 23.15 & 0.13 & -0.48&  0.15 & 0.86 & 0.17 & 0.48 & 0.16  &      \\
38 & 10.7 & $0 44 3.87$ & $41 18 16.2$ & 22.93 & 0.08 &  -   &  -   & -0.18 & 0.15 &  -   & -     &     \\
39 &  10.8 & $0 44 4.80$ & $41 18 31.0$ & 23.69 & 0.19 &  -   &  -   & 1.02 & 0.23  & -    & -     &     \\
40 & 10.9 & $0 44 4.12$ & $41 18 28.3$ &  -    & -    & -    & -   & 21.08 & 0.04  & 1.11 & 0.04  & 265895\\
41 & 11.1 & $0 44 5.22$ & $41 18 10.4$ &  -    & -    & -    & -   & 22.07 & 0.07  & 1.43 & 0.08  &      \\
42 & 11.1 & $0 44 5.62$ & $41 18 14.8$ &  -    & -    & -    & -   & 21.75 & 0.05  & 1.55 & 0.06  &      \\
43 & 11.3 & $0 44 4.92$ & $41 18 31.4$ & 22.32 & 0.08 & 0.87 & 0.13 &  0.70 & 0.09 & 1.13 & 0.06  &      \\
44 & 11.5 & $0 44 5.72$ & $41 18 16.3$ &  -    & -    & -    & -   & 21.82 & 0.07  & 2.38 & 0.07  & 265131\\
45 & 11.5 & $0 44 5.24$ & $41 18 10.1$ & 22.89 & 0.09 &  -   &  -  &  0.87 & 0.11  & 1.42 & 0.07  &      \\
46 & 11.5 & $0 44 4.46$ & $41 18 31.1$ & 21.84 & 0.05 & 0.69 & 0.08 & 0.81 & 0.07  & 1.39 & 0.05  & 266128\\
47 & 11.5 & $0 44 3.76$ & $41 18 22.3$ & 22.68 & 0.08 &  -   &  -   & 0.79 & 0.10  & 1.47 & 0.08  & 265573\\
48 & 11.7 & $0 44 5.66$ & $41 18 14.5$ & 23.04 & 0.10 & 0.62 & 0.20 & 1.32 & 0.11  & 1.54 & 0.05  &      \\
49 & 11.8 & $0 44 4.50$ & $41 18  8.8$ & 21.66 & 0.03 & 1.52 & 0.09 & 1.19 & 0.04  & 1.15 & 0.03  & 264615\\
50 & 11.9 & $0 44 3.98$ & $41 18 28.2$ & 22.65 & 0.14 & -0.26&  0.11 &   -  &  -   & 19.99 & 0.01 &       \\
51 & 12.1 & $0 44 5.50$ & $41 18 11.4$ & 22.43 & 0.06 & 0.61 & 0.12  & 0.54 & 0.10 & 1.01 & 0.09  &      \\
52 & 12.1 & $0 44 4.05$ & $41 18 11.1$ & 22.64 & 0.07 & 0.36 & 0.11  & 0.68 & 0.10 & 1.18 & 0.09  &      \\
53 & 12.2 & $0 44 5.67$ & $41 18 26.7$ & 20.58 & 0.03 & 0.18 & 0.03  & 0.15 & 0.03 & 0.43 & 0.04  & 265816\\
54 & 12.2 & $0 44 4.68$ & $41 18  8.1$ & 22.43 & 0.06 & 0.90 & 0.13  & 0.80 & 0.08 & 1.15 & 0.06  &      \\
55 & 12.3 &0 $44 5.34$ & $41 18 30.7$ & 22.90 & 0.15 &  -   &  -    & 0.76 & 0.16 &  -   & -     &     \\
56 & 12.3 & $0 44 4.44$ & $41 18 32.0$ &  -    & -    & -    & -   & 21.09 & 0.05  & 1.40 & 0.05  & 266128\\
57 & 12.4 & $0 44 5.86$ & $41 18 19.6$ & 22.39 & 0.08 & 1.03 & 0.15 & 0.84 & 0.10  & 0.14 & 0.11  & 265353\\
58 & 12.4 & $0 44 4.70$ & $41 18 32.6$ & 22.96 & 0.17 & 0.84 & 0.30 &  -   & -     & -    & -     &     \\
            \hline
         \end{array}
$$
\end{table*}
\setcounter{table}{2}
\begin{table*} 
\caption{continued}
$$
\begin{array}{l c c c c c c c c c c c c l}
\hline
$no.$ &  $d$(\arcmin) & \alpha_{2000} & \delta_{2000} &  $V$   & \sigma&  $B-V or B$ & \sigma &  $V-R  or R$ & \sigma & $R-I or  I$&
 \sigma &  $Mag.no.$  \\
\hline
59 & 12.6 & $0 44 3.66$ & $41 18 18.1$ &  -    & -    & -    & -   & 21.13 & 0.05 & -0.58 & 0.15  & 265294\\
60 & 12.7 & $0 44 5.14$ & $41 18 32.2$ & 22.68 & 0.10 & 1.19 & 0.20 & 0.39 & 0.13 & 1.12 & 0.11   &     \\
61 & 12.8 & $0 44 5.89$ & $41 18 19.9$ &  -    & -    & -    & -   & 21.52 & 0.11 & 0.14 & 0.11  & 265353\\
62 & 12.9 & $0 44 5.89$ & $41 18 22.2$ &  -    & -    & -    & -   & 21.89 & 0.06 & 1.21 & 0.07  & 265485\\
63 & 12.9 & $0 44 3.95$ & $41 18 11.2$ & 22.43 & 0.07 & 0.19 & 0.14 & 0.80 & 0.09 & 1.13 & 0.07  & 264754\\
64 & 13.0 & $0 44 5.68$ & $41 18 28.0$ & 22.72 & 0.10 & 0.48 & 0.15 & 0.29 & 0.16 & 1.23 & 0.17   &     \\
65 & 13.0 & $0 44 5.62$ & $41 18 11.5$ & 23.34 & 0.14 & 0.05 & 0.20 & 1.08 & 0.18 & 1.48 & 0.12   &     \\
66 & 13.7 & $0 44 5.89$ & $41 18 25.1$ & 22.73 & 0.13 & 0.60 & 0.18 & 0.86 & 0.14 & 1.40 & 0.06  & 265728\\
67 & 13.7 & $0 44 3.99$ & $41 18  9.5$ & 22.19 & 0.07 &  -   &  -   & 0.47 & 0.11 & 0.60 & 0.13  &      \\
68 & 13.8 & $0 44 5.90$ & $41 18 25.2$ &  -    & -    & -    & -   & 21.90 & 0.05 & 1.40 & 0.06  & 265728\\
69 & 13.9 & $0 44 5.60$ & $41 18 30.4$ & 22.67 & 0.11 & 1.46 & 0.25 & 0.46 & 0.14 & 1.43 & 0.11  &    \\  
70 & 14.0 & $0 44 5.99$ & $41 18 21.8$ & 22.53 & 0.08 &  -   &  -   & 0.50 & 0.12 & 1.12 & 0.10  &    \\  
71 & 14.1 & $0 44 3.95$ & $41 18  9.4$ &  -    & -    & -    & -   & 21.74 & 0.09 & 0.61 & 0.13  &    \\  
72 & 14.1 & $0 44 3.90$ & $41 18  9.9$ & 22.45 & 0.09 &  -   &  -  &  0.49 & 0.13 &  -   & -     &    \\ 
73 & 14.2 & $0 44 5.77$ & $41 18 11.7$ &  -    & -    & -    & -   & 22.10 & 0.10 & 0.99 & 0.10  &    \\  
74 & 14.3 & $0 44 4.54$ & $41 18  6.1$ & 22.24 & 0.08 & 0.20 & 0.10 & 1.00 & 0.09 & 1.30 & 0.06  &    \\  
75 & 14.4 & $0 44 3.49$ & $41 18 21.6$ & 22.03 & 0.08 &  -   &  -  &  0.98 & 0.09 & 1.16 & 0.03  &    \\  
76 & 14.5 & $0 44 6.04$ & $41 18 18.8$ & 22.37 & 0.06 & 1.10 & 0.22&  0.89 &0.08  & 0.76 & 0.07  &     \\ 
77 & 14.5 & $0 44 5.90$ & $41 18 26.8$ & 22.44 & 0.11 &  -   &  -  &  0.95 & 0.12 & 0.96 & 0.06  &    \\  
78 & 14.6 & $0 44 3.62$ & $41 18 13.4$ & 22.97 & 0.10 & 0.47 & 0.16&  1.55 &0.11 & 1.39 & 0.05   &     \\
79 & 14.6 & $0 44 6.05$ & $41 18 21.8$ & 22.67 & 0.11 &  -  &   -  &  0.64 & 0.13 & 1.12 & 0.12  &    \\  
80 & 14.6 & $0 44 5.34$ & $41 18 33.2$ & 21.78 & 0.04 & 1.18 & 0.08&  0.76 & 0.05 & 0.91 & 0.04  & 266223 \\
81 & 14.7 & $0 44 5.99$ & $41 18 15.3$ & 22.61 & 0.07 & 0.44&  0.11&  0.76 & 0.09 & 1.59 & 0.07  &      \\
82 & 14.7 & $0 44 3.50$ & $41 18 24.4$ & 21.75 & 0.05 &  - &    -  &  0.55 & 0.07 &  -   & -    & 265643 \\
83 & 14.8 & $0 44 3.59$ & $41 18 13.5$ &  -    & -    & -  &   -   & 21.86 & 0.07 & 0.77 & 0.09 &      \\ 
84 & 14.8 & $0 44 6.06$ & $41 18 18.2$ &  -    & -    & -  &   -   & 21.47 & 0.05 & 0.76 & 0.07 &       \\
85 & 14.8 & $0 44 4.46$ & $41 18  5.8$ & 21.73 & 0.06 &  - &    -  &  1.04 & 0.07 & 0.07 & 0.06  & 264388 \\
86 & 14.9 & $0 44 5.22$ & $41 18  6.3$ &  -    &  -   &  - &    -  & 21.70 & 0.07 & 1.19 & 0.07  & \\     
\hline
                 \end{array}
$$
        \end{table*}

\begin{acknowledgements}
We thank the following colleagues: R. Casalegno of INAF-Turin
Observatory for his help with the mosaic photometric software 
in the early stage of this project, G. Ivanov and P. Kunchev for providing us with the plate
material on M31, obtained with 2m telescope of the BAS, R. Supper
for the  information on the second {\sl ROSAT} survey of M31, C. Markwardt
for computer support.
M. Orio acknowledges funding of the University
of Wisconsin College of Letters and Sciences, of the Italian Space Agency 
(ASI) and a ``40\%'' 
grant of the Italian Ministry of University and Research for astrophysical
research on compact objects.
\end{acknowledgements}

\end{document}